\documentstyle[12pt,twoside,psfig]{article}
\newcommand{\beq}{\large\begin{equation}}
\newcommand{\eeq}{\end{equation}\normalsize}
\newcommand{\bma}{\large\[}
\newcommand{\ema}{\]\normalsize}

\def\bear{\begin{eqnarray}}
\def\ear{\end{eqnarray}}

\begin{document}

\pagestyle{empty}
\vspace*{2.0cm}

\centerline{\Large D.V.Skobeltsyn Institute of Nuclear Physics}
\centerline{\Large M.V.Lomonosov Moscow State University}

\vspace*{3.0cm}
\rightline{\large INP MSU Preprint 99--42/600}

\vspace*{2.0cm}
\centerline{\large Yu.A.~Golubkov{\small $^{a)}$},
M.Yu.~Khlopov{\small $^{b)}$}}

\vspace*{2.0cm}
\centerline{\LARGE Diffuse Gamma Flux from Antiproton Annihilation}
\centerline{\LARGE in Our Galaxy}

\vspace*{1.0cm}
\centerline{\it {\small $^{a)}$}Institute of Nuclear Physics,
Moscow State University, Vorobjevy Gory,}
\centerline{\it 119899, Moscow, Russia,}
\centerline{\it {\small $^{b)}$}Institute of Applied Mathematics,
Miusskaya Pl.4, 125047, Moscow, Russia,}

\vspace*{5.0cm}
\centerline{\large Moscow 1999}

\newpage
\vspace*{1.0cm}
\centerline{Yu.A.~Golubkov{\small $^{a)}$},
M.Yu.~Khlopov{\small $^{b)}$}}

\bigskip
\noindent $^{a)}$e-mail: golubkov@npi.msu.su\\
\noindent $^{b)}$e-mail: mkhlopov@orc.ru\\

\vspace*{1.0cm}
\centerline{\large Preprint of Institute of Nuclear Physics 99---42/600}
\vspace*{1.0cm}
\centerline{\Large Diffuse Gamma Flux from Antiproton Annihilation}
\centerline{\Large in our Galaxy}

\begin{abstract}%
The all-sky survey in high-energy gamma rays (E$>$30 MeV) 
carried out by the Energetic Gamma Ray Experiment Telescope (EGRET)
aboard the Compton Gamma-Ray Observatory provides a unique  
opportunity to examine in detail the diffuse gamma-ray 
emission. The observed diffuse 
emission has a Galactic component arising from cosmic-ray 
interactions with the local interstellar gas and radiation as 
well an almost uniformly distributed component that is 
generally believed to originate outside the Galaxy. 
The results of the observations have been interpreted as the 
extragalactic high-energy gamma-ray emission arising
primarily from unresolved gamma-ray-emitting blazars.

Here we consider another possible origin
of the diffuse gamma--ray flux, namely, as originating from
the annihilation of the antiprotons with the interstellar medium.
\end{abstract}

\vspace*{7.0cm}
\leftline{\small{\copyright Institute of Nuclear Physics,
Moscow State University, 1999}}

\newpage
\pagestyle{plain}


%

{\footnotetext[1]{Invited talk at the IV Int. Conf. ''Cosmion-99'',
Moscow, October 17-24, 1999}}
\section{The annihilation cross sections}

The existing theoretical models based mainly on the partonic picture of the hadronic interactions are definitely invalid for $\bar pp$ annihilation at low energies. Due to this fact we used experimental data as for evaluation of the annihilation cross section as well as for the simulation of the final state and secondaries distribution.

The annihilation cross section is the difference between total and inelastic
ones $\sigma_{ann}\,\approx\,\sigma_{tot}-\sigma_{el}$.
As it follows from data the dependence 
$\sigma_{ann}\,\sim\,v^{-1}$ is valid already for laboratory antiproton
momenta $p_{lab}\,\le\,1000$ MeV/c.
Thus, at $P_{lab}\,\ge\,300\ MeV/c$ we used data from \cite{PDG96} 
for the total and elastic cross sections and
at momenta lesser than $300$ MeV we used the dependence

\bma
\sigma_{ann}(P\,<\,300\ \mbox{MeV}) \ = \ \sigma_0\ C(v^*)/v^*
\ \  {\mbox{for annihilation cross section}}\,,
\ema

\noindent and

\bma
\sigma_{el}\ =\ const\ \ {\mbox{for elastic cross section}}
\ema

The additional Coulomb factor $ C(v^*)$ gives large increase 
for the annihilation cross section at small velocities of the antiproton
and is defined by the expression \cite{Landau3}:

\bma
C(v^*)\ =\ \frac{2\,\pi\,v_c/v^*}
{1\,-\,\exp\left ( -2\,\pi\,v_c/v^*\right )},
\ema

\noindent where, $v_c\,=\,\alpha\,c$, with $\alpha$ being the fine structure
constant and $c$ being the speed of light and $v^*$ is the velocity
of the antiproton in the center-of-mass $\bar pp$ system.

We used the experimental data on the $\bar pp$ annihilation cross section \cite{Bertin96,Bruckner90} to normalize the $1/v$ behaviour and the Coulomb factor, choosing:

\bma
\sigma_0 \ = \ \sigma_{ann}^{exp}(P\,=\,300\ \mbox{MeV})
\ \approx\ 160\ \mbox{mb}\,.
\ema

\section{Gamma flux}

We used the experimental data \cite{Backenstoss83} on the $\bar pp$
annihilation at rest to simulate the distribution of the final state particles,
see Table.


\vspace*{0.5cm}

\vbox{
\centerline{Table. Relative probabilities of $\bar pp$ annihilation channels.}
\bma
\begin{array}{|l|r||l|r|}
\hline
&&&\\
\mbox{\rm Channel} & \mbox{\rm Rel. prob.}, \% 
& \mbox{\rm Channel} & \mbox{\rm Rel. prob.}, \%\\
&&&\\
\hline
&&&\\
\pi^+\pi^-\pi^0   & 3.70     & 2\,\pi^+2\pi^-\eta  & 0.60\\
\rho^-\pi^+       & 1.35     & \pi^0\rho^0 & 1.40 \\
\rho^+\pi^-       & 1.35     & \eta\rho^0 & 0.22\\
\pi^+\pi^-2\pi^0 & 9.30      & 4.99\,\pi^0  & 3.20\\
\pi^+\pi^-3\pi^0 & 23.30     & \pi^+\pi^- & 0.40\\
\pi^+\pi^-4\pi^0 & 2.80      & 2\,\pi^+2\pi^- & 6.90\\
\omega\pi^+\pi^- & 3.80      & 3\,\pi^+3\pi^- & 2.10\\
\rho^0\pi^0\pi^+\pi^- & 7.30 & K\bar K \,0.95\pi^0 & 6.82\\
\rho^+\pi^-\pi^+\pi^- & 3.20 & \pi^0\eta ' & 0.30\\
\rho^-\pi^+\pi^+\pi^- & 3.20 & \pi^0\omega & 3.45\\
2\,\pi^+2\pi^-2\pi^0 & 16.60 & \pi^0\eta & 0.84\\
2\,\pi^+2\pi^-3\pi^0 & 4.20  & \pi^0\gamma & 0.015\\
3\,\pi^+3\pi^-\pi^0 & 1.30   & \pi^0\pi^0 & 0.06\\
\pi^+\pi^-\eta  & 1.20 &&\\
&&&\\
\hline
\end{array}
\ema
}

In practice, the approximation of the annihilation at rest is valid with very good accuracy up to laboratory momentum of the incoming antiprotons about $0.5$ GeV because at this laboratory momentum the kinetic energy of the antiproton is still order of magnitude lesser than the twice antiproton mass.
The average number of $\gamma$'s per annihilation is

\bma
N_{\gamma}\ =\ 3.93\,\pm\,0.24\,.
\ema

The simulation of the distributions of final state particles 
has been performed according to phase space in the center-of-mass of the 
$\bar pp$ system. After that PYTHIA 6.127 package \cite{Pythia6} has been used to perform the decays of all unstable and quasi--stable particles.

We used the spherical model for halo and parametrized the density distribution
of interstellar hydrogen gas along the $z$ direction (to North Pole) as:

\bma
n_H(z)\ =\ n_H^{halo}\,+\,\frac{n_H^{disc}}{1\,+\,(z/D)^2},
\ema

\noindent where $n_H^{halo}\,=\,10^{-2}$ cm$^{-3}$ is the hydrogen
number density in the halo, $n_H^{disc}\,=\,1$ cm$^{-3}$ is the hydrogen
number density in the disc and $D\,=\,100$ pc is the half-width of the
gaseous disc.
The number of gammas arriving from the given direction is defined
by the well known expression:

\bma
J_{\gamma}(E)\ =\ \int_0^L\,dl\,\psi (E,r,z).
\ema

Here $\psi (E)$ is the density of gamma source along the observation direction
$l$ in suggestion of the isotropic distribution of the gamma source. 
The integration must be performed 
up to the edge of the halo $L\,=\,-\alpha_x\,R_{\odot}\,+\,
\sqrt{R_{halo}^2-R_{\odot}^2\,\left (1-\alpha_x^2\right )}$
with $\alpha_x$ being the cosine of the line-of-sight to the $x$ axis.

The flux of the antiprotons has been chosen similar to the standard
proton flux:

\bma
J_{\bar p}(E_{kin})\ =\ N_{tot}\,\beta
\,\left (\frac{1\,\mbox{GeV}}{E_{kin}}\right )^{2.7},
\ema

\noindent where $\beta\,=\,p/E$ is the dimensionless antiproton velocity. 
The minimal kinetic energy of the antiprotons has been chosen equal to the velocity dispersion in the Galaxy, $v_{min}\,\approx\,300$ km/s, i.e.,

\bma
E_{kin}^{min}\ \approx\ 500\ \mbox{eV}.
\ema

\begin{figure}[htb]                
\par
\centerline{\hbox{%
\psfig{figure=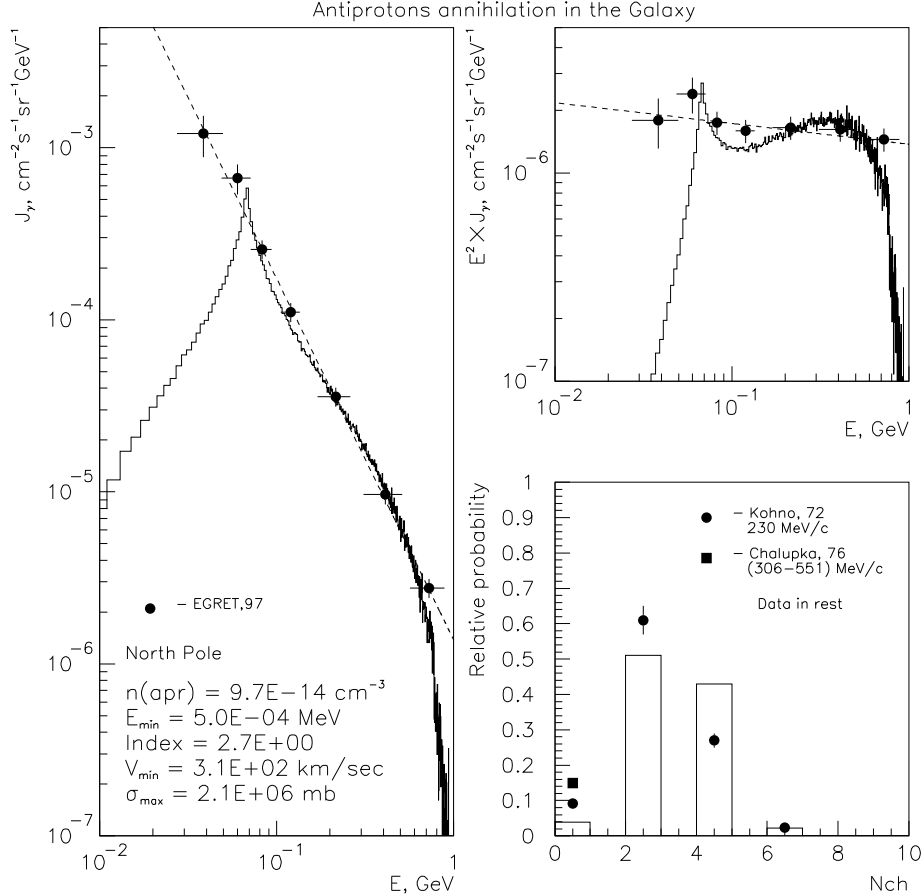,bbllx=1.5cm,bblly=5.5cm,%
bburx=19.0cm,bbury=23.0cm,clip=t,height=12.0cm}%
}}
\par
\caption{\label{gmflux}
Comparison of the calculated differential fluxes 
of $\gamma$ quanta from $\bar pp$
annihilation with experimental data {\em EGRET} \cite{EGRET97}
on diffuse gamma background (a,b). There is also shown the charged
multiplicity distribution in the annihilation model described in the text (c).
}
\end{figure}

Fig.\ref{gmflux}(a,b) demonstrates the resulting differential gamma 
distribution in the Galactic North Pole direction
in comparison with EGRET data \cite{EGRET97} in the range $10\,\le\,E_{\gamma}\,\le\,1000$ MeV.
The peak of $\pi^0$ decay is clearly seen both in calculations as well
as in experimental distributions. Fig.\ref{gmflux}(c) shows the charged
multiplicity distribution in the annihilation model described above.
Experimental points have been taken from \cite{Kohno72,Chaloupka76}.

At this choice the necessary integral number density of the antiprotons
$N_{tot}$ is:

\bma
N_{tot}\ =\ \int_{E_{min}}^{\infty}\,\frac{dN_{\bar p}(E)}{dE_{kin}}
\ \sim\ 10^{-13}\ \mbox{cm}^{-3}.
\ema

Note here, that within reasonable variation of the index of the
antiproton flux the integral number of the antiprotons does not varies significantly. This is due to the fact that the main contribution
comes from the $\bar pp$ annihilation at the lowest velocity of the antiprotons.
If we choose the minimal velocity of antiprotons according the
velocity of the stellar wind, $v_{wind}\,\approx\,1000$ km/s,
the necessary antiproton number density must be increased by order
of magnitude and is $N_{tot}\,\sim\,10^{-12}$ cm$^{-3}$. This number
also does not contradict to any theoretical or experimental limits
on relative amount of the antimatter.

\section{Discussion. Antimatter Globular Cluster}

We considered the stationary case, provided by the presence of
a permanent source of the antimatter.
Assume \cite{Khlopov98} that antimatter stars in the Galaxy form globular
cluster, moving along elliptical orbit in the halo. In the first approximation
the integral effect we study depends on the total mass of the
antimatter stars in the Galaxy and it does not depend on the amount of globular clusters.
The main contribution into galactic gamma radiation comes from
the antimatter lost by the cluster and spread over the Galaxy.

There are two main mechanisms of antimatter loss by the cluster. The first one
is the stationary mass loss by antimatter stars in the form of stellar wind.
The second mechanism is the antimatter Supernova explosions.
In both cases the antimatter is spread over the Galaxy in the form of positrons
and antinuclei. Due to traveling in the halo magnetic fields the emitted antiprotons fill the whole halo and one can consider their number density to be constant over the halo.

\begin{figure}[htb]                
\par
\centerline{\hbox{%
\psfig{figure=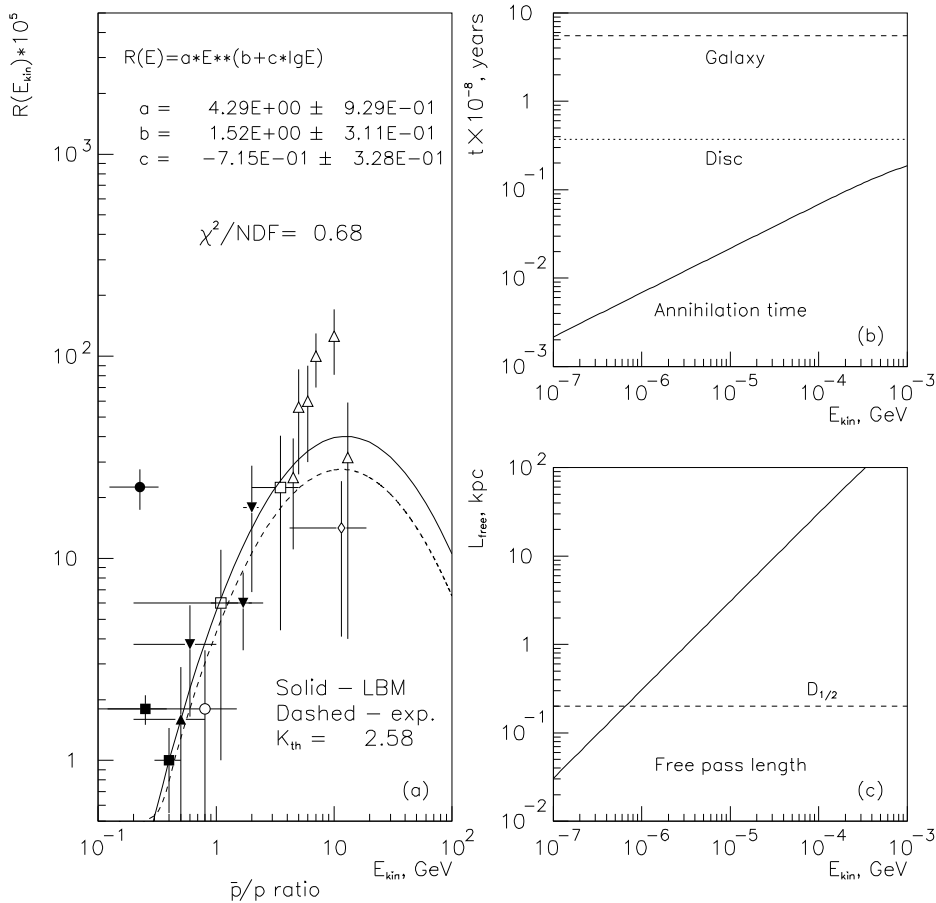,bbllx=1.5cm,bblly=5.5cm,%
bburx=19.0cm,bbury=23.0cm,clip=t,height=12.0cm}%
}}
\par
\caption{\label{expfit}
(a) Fit of the $\bar pp$ ratio to experimental data.
Solid line shows predictions of the two--zone leaky box model, increased by factor $K\,\approx\,2.6$. Dashed curve is the phenomenological fit, described in the text.
(b) The comparison of the antiprotons annihilation time in the disc (solid)
with the predicted LBM confinement time from Galaxy (dashed) and from disc (dot-dashed). (c) The free pass length of antiprotons in the disc.
Dashed line shows the half-width of the disc.
}
\end{figure}

To evaluate the confinement time for the antiprotons in the Galaxy we used the results of the ''two--zone'' leaky box model \cite{Chardonnet96}.
As the confinement time enters as a common factor in the predicted $\bar pp$ ratio \cite{Chardonnet96}, we can find the necessary factor, performing the
fit of the normalization to the experimental data.
We removed from the fit two the most left points, strongly affected by the heliosphere \cite{Geer98}. Solid curve in Fig.\ref{expfit}(a) presents
the LBM predictions for the $\bar pp$ ratio,
multiplied by the fitted factor $K\ =\ 2.58$, which factor increases the confinement time for slow antiprotons in the Galaxy up to $2\cdot10^8$ years.   
Dashed curve is the phenomenological fit in the form $R(E)\,=\,a\,E^{b+c\lg E}$.
Experimental points have been taken from \cite{Golub98} where references on the data can be found.

Fig.\ref{expfit} shows also 
the time of life of the antiprotons in the disc (b) and the free pass length
of the antiprotons in the disc (c) versus antiproton kinetic energy.
The correspondent curves for halo can be easily obtained if one reduces
the hydrogen number density by two-three orders of magnitude
according to $H$ number density in the halo.
Note here that the presented in Fig.\ref{expfit} confinement time for very slow antiprotons has been extrapolated to small kinetic energies, suggesting that
$t_{esc}(E_{kin}<0.1\,GeV)\ \approx\ const\ =\ t_{esc}(E_{kin}=0.1\,GeV)$
and is, in fact, the lower limit for the confinement time.

From Fig.\ref{expfit}(b) we can conclude that confinement time for low-energy antiprotons in the halo and in the gaseous disc is much greater than their life time with respect to the annihilation. Therefore we can consider the picture as a stationary one. From Fig.\ref{expfit}(c) we see, that free pass length of the slowest antiprotons is lesser than the half-width of the disc and we can observe the products of the annihilation only but not the original antiprotons. 
The detailed analysis of the antimatter annihilation mechanism for Galactic diffuse gamma flux in the terms of antimatter globular cluster and the estimation of the total amount of antimatter stars and the expected antihelium flux for this case will be considered in the successive paper.

%
%

\bigskip
{\it Acknowledgements}.
The authors acknowledge the COSMION Seminar participants for useful discussions.
The work was partially supported by Section ''Cosmoparticle Physics'' of State Scientific Technical Programme ''Astronomy. Fundamental Space Research'' and performed in the framework of COSMION--ETHZ and AMS--EPICOS collaborations of ASTRODAMUS project.


\end{document}